\documentclass[english,preprint,showpacs,pra,superscriptaddress]{revtex4-1}
\usepackage{graphicx}
\usepackage{amsfonts,amsmath,amssymb}
\usepackage{enumerate}
\usepackage{epsfig}
\usepackage{color}
\usepackage{amscd}

\usepackage{epstopdf}

\providecommand{\tabularnewline}{\\}

\begin{document}

\title{Measurement-device-independent quantum key distribution with nitrogen
vacancy centers in diamond}

\author{Nicol$\acute{\rm \bf o}$ Lo Piparo}
\affiliation{School of Electronic and Electrical Engineering, University of Leeds, Leeds, UK}
\affiliation{National Institute of Informatics, 2-1-2 Hitotsubashi, Chiyoda, Tokyo 101-0003, Japan.}
\author{Mohsen Razavi}
\affiliation{School of Electronic and Electrical Engineering, University of Leeds, Leeds, UK}
\author{William J. Munro}
\affiliation{National Institute of Informatics, 2-1-2 Hitotsubashi, Chiyoda, Tokyo 101-0003, Japan.}
\affiliation{NTT Basic Research Laboratories, NTT Corporation, 3-1 Morinosato-Wakamiya, Atsugi, Kanagawa, 243-0198, Japan.}







\begin{abstract}
Memory-assisted measurement-device-independent quantum key distribution (MA-MDI-QKD) has recently been proposed as a possible intermediate step towards the realization of quantum repeaters. Despite its relaxing some of the requirements on quantum memories, the choice of memory in relation to the layout of the setup and the protocol has a stark effect on our ability to beat existing no-memory systems. Here, we investigate the suitability of nitrogen vacancy (NV) centers, as quantum memories, in MA-MDI-QKD. We particularly show that {\em moderate} cavity enhancement is required for NV centers if we want to outperform no-memory QKD systems.  Using system parameters mostly achievable by the today's state of the art, we then anticipate some total key rate advantage in the distance range between 300~km and 500~km for cavity-enhanced NV centers. Our analysis accounts for major sources of error including the dark current, the channel loss, and the decoherence of the quantum memories.
\end{abstract}
\maketitle

\section{Introduction}

Long-distance quantum key distribution (QKD) should ideally enable
the exchange of secret data without the need to trust intermediate
nodes \cite{IEEE2}. Quantum repeaters \cite{Briegel1998,dur1999,sangouard2011,munro2015} 
are often considered to be the main means
to achieve this goal, but they are facing numerous technological challenges,
e.g., the development of reliable quantum memory (QM) units, which delay
their implementation. A potentially feasible approach to increase
the quantum communication range has been proposed in \cite{abruzzo,panayi},
where the authors introduce memory-assisted measurement-device-independent
QKD (MA-MDI-QKD) schemes. Their protocols resemble a single-node quantum
repeater link with QMs only in the middle node, and optical encoders only at the users
end. Alternatively, one can look at them as MDI-QKD links \cite{MDI1,MDIQKD-404km,PRF-MDIQKD}, with additional QMs in the middle. The performance of these memory-assisted schemes much relies
on their employed QMs. Initially, ensemble-based memories were considered
as suitable candidates for such systems because of their short sub-nanosecond
writing times \cite{Walmsley:PRL:2010}. It turned out, however, that,
within the proposed schemes in \cite{panayi}, the multiple-excitation
effect in such QMs would prevent the MA-MDI-QKD protocol to beat the
no-memory QKD schemes \cite{IEEE1}. To avoid such problems, in this
paper, we investigate the suitability of nitrogen vacancy
(NV) centers in diamonds as quantum memories and show that it is
possible to beat the existing no-memory schemes, in terms of rate
versus distance, in certain regimes of interest. 


There are several possible solutions by which we can avoid the multiple-excitation effect in ensemble-based memories. In one approach, as proposed in \cite{IEEE1}, one can attempt to locally generate a pair of entangled photons, and then try to load the QM with one of them. If each photon in the entangled pair is truly a single photon, we would, in principle, excite only one atom in the ensemble. Another solution is to use single-atom/ion or quasi-single-atom QMs, such as quantum dots and NV centers in diamond. Each of these solutions would offer certain advantages and disadvantages, and while none could necessarily offer a practical advantage at this very time, it would be interesting to see how far each technology is from beating a no-QM system.

{Among various candidates for the QM, in this paper, we focus on the potential of NV centers in diamond. There is some evidence that such systems might offer better performance than their rivals, while a rigorous analysis in each case is needed to find out what would be the best each system can offer. For instance, in the case of quantum dots, one possible drawback could be their 
often very short spin coherence time, $T_{2},$ ranging from 2~ns to over
200~ns \cite{QD1,QD2}, which could prove too short to be effective
in the MA-MDI-QKD setup, {as we will show in Sec.~\ref{sec:results}}. The electronic spins in NV centers, instead, have coherence times on the order of milliseconds, which can be extended to seconds when their electron
spin state is transferred to nuclear spins \cite{Nuclear_spin_time2,dep_pap,Nuclear_spin_time}. As compared to single atoms and ions, NV centers offer faster interaction times with photons, on the order of tens of nanoseconds, while the former are generally slower systems. Since the short access time is one of the requirements in \cite{panayi}, the NV centers could then have an advantage in this regard as we numerically compare these systems in Sec.~\ref{sec:results}. 
NV centers, nevertheless, similar to any other single-atom-like QM, must be embedded into cavities if efficient
coupling with photons is required.}

{One of the key requirements in some of the protocols proposed in \cite{panayi} is the ability to entangle QMs with photonic states. In order to achieve a high key generation rate that can beat conventional no-QM systems, this entangling procedure 1) must have a reasonably high rate of success on the order of $0.1$; 2) must be repeatable with a rate roughly exceeding 10 MHz; and 3) must offer a high-fidelity (low error) operation. There have been various attempts in the field to entangle NV centers with single photons. In early experiments, both fidelity and the success rate are often low. For instance, in \cite{NVexp1}, the probability of creating spin-photon entanglement is on the order of $10^{-6}$, which is extremely low for the application we have in mind. The achieved fidelity is also rather low at around $70$\% \cite{NVexp1}. More recent experiments improve the fidelity, but the success rate still remains at a similar level \cite{NVexp2}. One key reason for the latter is the low collection efficiency of the photons coming out of the NV center. The efficiency would increase if instead of generating a photon entangled with the NV center, we first generate entangled photons and then store one of the photons in the NV center. An overall efficiency of 20\% has been reported in \cite{Wrachtrup} for transferring the state of a single photon to the nuclear spin of an NV center. But, then, such a system needs to be driven by a high-rate source in order to compete with no-QM systems that can be driven at GHz rates. It follows then, both for boosting the coupling efficiency and/or generating spectrally matched single photons at a high rate, we need to embed the NV centers in compact optical cavities.} 


{NV centers, embedded into cavities, can in principle satisfy all the requirements in MA-MDI-QKD. In \cite{Bill_paper}, the authors propose an innovative scheme for {\em cavity-enhanced}
NV centers that can potentially create memory-photon entangled states
with an extremely high fidelity ($F>0.99$) and high entangling rates. There,
the authors use two NV centers in diamond, each inside a cavity, to
create a spin entangled pair. The essence of this method is based
on how the NV center state affects the reflectivity of the cavity
system \cite{duan_kible}. In our work, we will modify the scheme in \cite{Bill_paper} to create spin-photon entanglement between the electron spin of an NV center, embedded into a cavity, and a single photon.} 


{Our main contribution is a rigorous and quantitative assessment of the applicability of NV centers in MA-MDI-QKD setups. We start with reviewing the experimental setups that couple single photons and NV centers and overestimate their performance in the context of MA-MDI-QKD. It turns out that none of these setups is capable of beating conventional QKD systems. Our key proposed solution is then based on cavity-enhanced NV centers. We show that even with some moderate cavity enhancement the rate-versus-distance behavior can substantially improve}. While the fabrication and testing of such devices is underway, we use the meticulous analysis in \cite{Bill_paper} to estimate the potential of such QMs in our setup.  We calculate the secret key generation rate, as the main figure of merit, for a number of NV-center-based MA-MDI-QKD schemes, and compare it with that of the no-memory system, as well as other main single-excitation candidates for QMs. Our analysis accounts for major sources of imperfection such as dark current in detectors and path loss as well as the decoherence of the QMs.



The paper is structured as follows. In Sec.~\ref{Sec:Basics}, we review the MA-MDI-QKD schemes proposed in \cite{panayi,IEEE1}, highlighting their key features and updating their measurement procedures. In Sec.~\ref{Sec:withNV}, we investigate the applicability of the non-cavity schemes proposed in \cite{NVexp1,NVexp2, Wrachtrup} for MA-MDI-QKD and propose memory-assisted schemes that use cavity-enhanced NV centers as memories. In Sec.~\ref{sec:results}, we describe our methodology for calculating the secret key generation rate for the proposed protocols. We continue by providing some numerical results before we draw our conclusions in Sec.~\ref{Sec:Conc}.

\section{Memory-assisted MDI-QKD: The basics}
\label{Sec:Basics}

\begin{figure}
\begin{centering}
\includegraphics[scale=0.9]{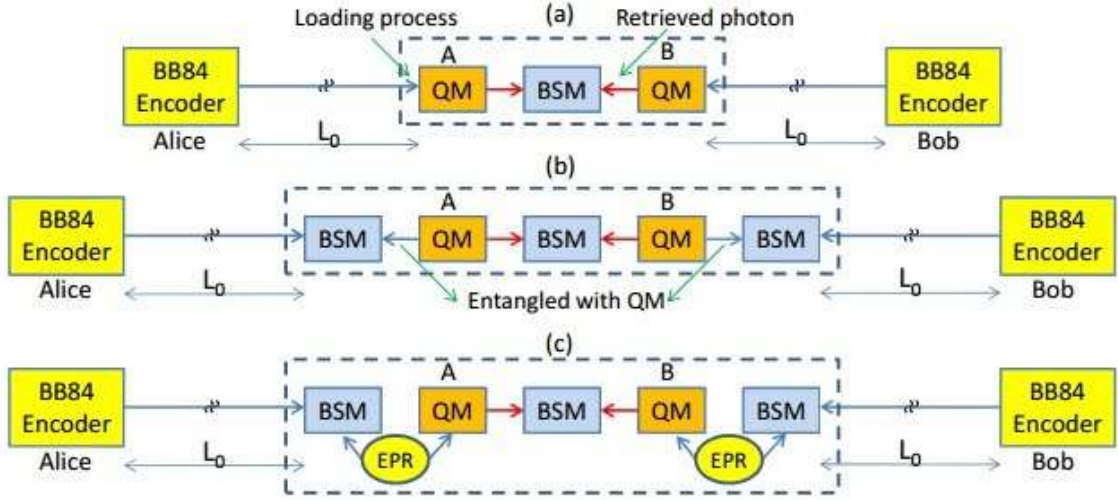} 
\par\end{centering}
\protect\protect\protect\caption{\label{fig:mem_assisted_setups}Different setups for memory-assisted
MDI-QKD, as proposed in \cite{panayi,IEEE1}, for (a) directly heralding and (b,c) indirectly heralding quantum memories. In (c), the EPR source generates an entangled pair of photons, but the photon will be written into the memory only if the side-BSM is successful (delayed writing). }
\end{figure}

MA-MDI-QKD can be implemented in different ways using different quantum memory modules. The original schemes proposed in \cite{panayi} were divided into two categories of directly versus indirectly heralding schemes; see Figs.~\ref{fig:mem_assisted_setups}(a) and (b). Later, in \cite{IEEE1}, the authors proposed a third setup using EPR sources; see Figs.~\ref{fig:mem_assisted_setups}(c). In all these setups, one needs to store the state of an incoming BB84-encoded photon, in a heralded way, into the QM. Once both memories are loaded, we need to perform a Bell-state measurement (BSM) on the QMs' states to generate, using the time-reversed entanglement idea \cite{biham}, correlated data between Alice and Bob \cite{MDI1}. The fact that this BSM is only done once we know of the storage of the transmitted photons is the key to improving the rate-versus-distance behaviour, as now the rate would, in principle, scale with the loss over half of the channel.

Depending on the MA-MDI-QKD scheme used, there are different requirements that need to be met. In Fig.~\ref{fig:mem_assisted_setups}(a), where, for each transmitted photon, we attempt to store it into the QM, we need to be able to verify whether or not the photon's state has successfully been captured by the QM. In such a protocol, the time period at which the whole loading scheme can be repeated cannot be shorter than the sum of three key time parameters: the {\em interaction time} between a photon and a QM needed to transfer the state between them, the {\em verification time} needed to establish if the loading has been successful, and the {\em preparation/initialization time} required to prepare the system back into a state that can interact with the next arriving photon. The repetition rate is an important factor for MA-MDI-QKD because in order to have a chance at beating no-QM systems, typically driven at GHz rates, we cannot afford to have slow memories. Another issue with low repetition rates is the requirement for longer coherence times. In \cite{panayi}, the authors show that for MA-MDI-QKD in Fig.~\ref{fig:mem_assisted_setups}(a) to have a chance at beating no-QM systems, one needs repetition rates exceeding 10~MHz and coherence times roughly 10000 times longer than the repetition period. For that reason, typical candidates with directly heralding features, such as trapped atoms/ions, may not perform their best within the setup of Fig.~\ref{fig:mem_assisted_setups}(a). For instance, in \cite{Rempe}, the authors use $^{87}\mathrm{Rb}$ atoms to realize the heralded transfer of a polarization qubit from a photon
onto a single atom. However, the initialization time of the atom is around 140 $\mu\mathrm{s}$, which restricts the repetition rate to below 10~kHz. 

In order to have the option of using other types of memories, in the schemes of Figs.~\ref{fig:mem_assisted_setups}(b) and (c), the photon storage is heralded in an indirect way by {\em teleporting} the user's photon into the QM. In order to do so, we first need to entangle a photon with the QM, and then do an additional side-BSM on this photon and the one sent by the user. A successful side-BSM heralds the storage of the photon. In Fig.~\ref{fig:mem_assisted_setups}(b), the entangled photon is generated by manipulating the QM. In that sense, the repetition period is determined by similar time parameters as before, except that now the interaction time refers to the time that it takes to entangle a photon with a QM after the initialization phase. The verification time in this case is effectively the time for doing the side-BSM, which can be very short. The same requirements are then held as in the scheme of Fig.~\ref{fig:mem_assisted_setups}(a) regarding the short repetition times and large storage-bandwidth products. In many QM setups, the former can be hard to achieve especially if cooling is required for the QM. For slower QMs, but the ones with long coherence times, one can then use the scheme in Fig.~\ref{fig:mem_assisted_setups}(c), in which we only manipulate the QM if we have a successful side-BSM. In this scheme, we can run the system at the rate at which entangled photons can be generated by the EPR source. Once we have a successful side-BSM, we trigger the writing procedure for storing the unused photon of the EPR source into the QM. In short distances, the rate will be cropped by the slow rate of the QM's preparation time, but, at long distances, we can effectively prepare the QM before the next photon survives the path loss. Using this trick, we can achieve a higher rate from slow QMs.

\begin{figure}
\begin{centering}
\includegraphics[scale=0.7]{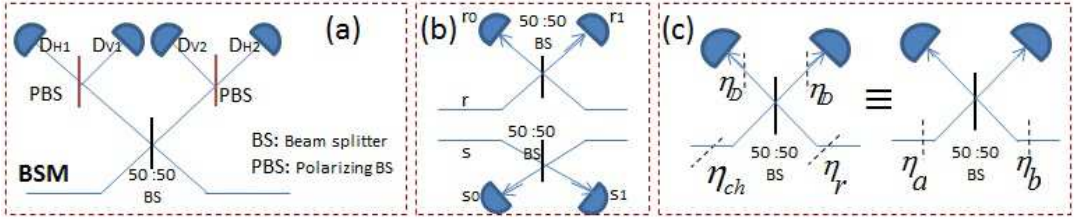}
\par\end{centering}

\protect\caption{\label{fig:BSM_pol}Bell-state measurement modules for (a) polarization
and (b) phase-encoded states. (c) The common building block in (a)
and (b), and its simplified version (on the right), when the setup's
inefficiencies are considered. In (c), $\eta_{r}$, $\eta_{D}$, and
$\eta_{ch}$, respectively, represent the reading, detector, and channel
efficiencies, and $\eta_{a}$ $=$ $\eta_{D}\eta_{ch}$ and $\eta_{b}$
$=\eta_{D}\eta_{r}$ .}
\end{figure}

Finally, to generate a raw key bit, one needs to do the middle BSM in Fig.~\ref{fig:mem_assisted_setups}. In \cite{panayi}, the authors assume that the states of the QMs can be transferred back to photons (the QMs carry no information from that point on), and then we can use the type of linear optical modules shown in Figs.~\ref{fig:BSM_pol}(a) and (b) to perform a partial BSM. In this paper, we refer to this scheme by the \textit{reading} protocol. {As we will discuss in the next section, for NV centers, this final BSM is not without its own challenges. In particular, in order to use the reading protocol, we need to find a double-$\Lambda$ structure in NV centers with identical energy gaps. This turns out to be nontrivial for NV centers}. An alternative approach is to again entangle a photon with each QM and do the partial BSM on these photons. This is known as the \textit{double-encoding} scheme \cite{RUS}. If the BSM is successful, a further $X$-basis measurement needs to be done on the QMs to enable an indirect BSM on the memories' states. The double-encoding technique also turns out to be not feasible or quite inefficient for many existing spin-photon entangling schemes that rely on NV centers. One way to make this scheme more efficient is to use cavity-enhanced NV centers, as we consider in this paper. Finally, one can potentially use a direct BSM on QMs without any interaction with photons. In the case of NV centers, this can be done if we use both the nuclear and electron spins in a single NV center \cite{QCMC2016}. We will investigate such an option in a separate work. Note that, in Figs.~\ref{fig:mem_assisted_setups}(b) and (c), side BSMs are performed on two optical modes, for which we use the linear optical module in Fig. \ref{fig:BSM_pol}(a), for polarization encoding, or the one in Fig. \ref{fig:BSM_pol}(b) for phase encoding systems.

\section{Memory-assisted MDI-QKD with NV centers}
\label{Sec:withNV}

In this section we consider several avenues for employing NV centers, as QMs, in any of the setups in Fig.~\ref{fig:mem_assisted_setups}. This can be divided into two categories: 1) experiments in which an NV center has been entangled with a photon, or a single photon has been written into the memory. The common feature in these experiments is that in none of which the NV center is embedded into a microcavity; and 2) the proposed setups for cavity-enhanced NV centers, which, while not yet being experimentally demonstrated, we have sufficiently rigorous analytical results to estimate their performance. One of our key findings in this paper is that none of the main candidates in the first group is capable of beating no-QM systems for two fundamental reasons. First, the entangling/collection efficiency is often very low in such experiments when there is no confining cavity around the NV center. Secondly, in most experiments, there is no straightforward way to perform the middle BSM by either reading or double-encoding protocol. Both issues can be rectified if we use cavity-enhanced setups as we show in this section.

\subsection{MDI-QKD with non-cavity NV centers}

There is a range of experiments on spin-photon interactions in NV centers. Here, we consider three representative examples and explore whether they can offer any advantages in the context of MA-MDI-QKD. The first of such is the early experiment reported in \cite{NVexp1}, followed by more recent experiments by Hanson's group \cite{NVexp2, Pfaff532}. The last example is about efficient transfer of single photons into NV centers reported in \cite{Wrachtrup}. For none of these setups, however, we were able to find or come up with an efficient readout scheme as required for the final BSM operation. In the lack of a proper working scheme, we introduce a toy model to estimate what at best non-cavity systems can offer.

\begin{figure}
\begin{centering}
\includegraphics[scale=0.8]{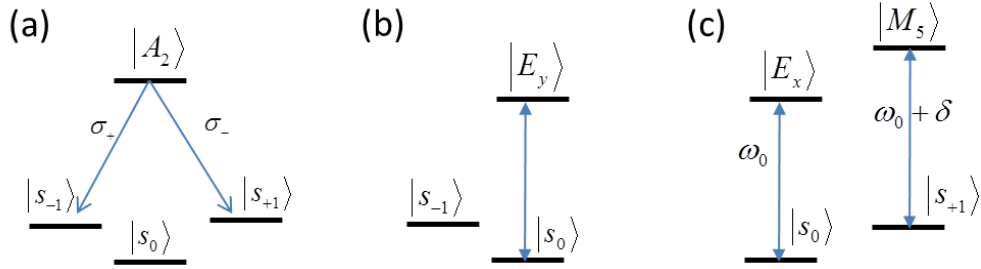}
\par\end{centering}

\protect\caption{\label{fig:scheme1_en_lev} The relevant energy level structure for the NV center used in (a) \cite{NVexp1}, (b) \cite{NVexp2}, and (c) \cite{Bill_paper}.}

\end{figure}
 
The first candidate we consider for spin-photon entanglement in Fig.~\ref{fig:mem_assisted_setups}(b) is the setup proposed in \cite{NVexp1}. In this setup, see Fig.~\ref{fig:scheme1_en_lev}(a), the NV center is prepared in a specific excited state $|A_{2}\left\rangle \right.$ that ideally decays with equal probability into two different long lived spin states, namely, $|s_{\pm1}\left\rangle \right.$, representing the $\pm1$ electron spin states. Such a transition would correspondingly result in emitting orthogonal circularly polarized photons, $|\sigma_{+}\left\rangle \right.$ and $|\sigma_{-}\left\rangle \right.$, in the following entangled state with the QM \cite{NVexp1}
\begin{equation}
|\Psi_{1}\left\rangle \right.=\frac{1}{\sqrt{2}}\left(|\sigma_{-}\left\rangle |0\left\rangle_{\rm NV} \right.\right.+|\sigma_{+}\left\rangle |1\left\rangle_{\rm NV} \right.\right.\right),\label{eq:NV_center1}
\end{equation}
where $|0\rangle_{\rm NV} = |s_{+1}\rangle$ and $|1\rangle_{\rm NV} = |s_{-1}\rangle$.

There are several practical issues with the above entangling procedure. { First, there is the issue of efficiency. In this setup, a combination of  weak NV-center-photon coupling and poor collection efficiency results in a very low success rate on the order of $10^{-6}$ \cite{NVexp1}. Furthermore, only a small fraction of photons are emitted into the zero-phonon line, while most are emitted into the phonon sidebands, where the latter would not result in the required spin-photon entanglement \cite{NV1}}. The latter is mainly responsible for the rather low fidelity of this scheme at around 70\%. Finally, once both QMs in Fig. \ref{fig:mem_assisted_setups}(b) are loaded, we need to somehow perform the central BSM operation on these QMs. This is, however, a challenging task within this setup, as neither the reading nor the double-encoding scheme can easily be implemented in this setup.

Some of the problems with the scheme in \cite{NVexp1} can be rectified by the scheme proposed in \cite{NVexp2}. In particular, here, the authors use resonant versus non-resonant transitions to have a conditional single photon generation. In Fig.~\ref{fig:scheme1_en_lev}(b), if the NV center is in $|s_0\rangle$, a resonant transition to state $|E_y\rangle$ would result in a spontaneous photon transmission, whereas, for an NV center in $|s_{-1}\rangle$, we do not expect any photons emerging. This process would ideally result in an entanglement between the number of photons, zero or one, in the collected photonic mode and the subspace spanned by $|s_0\rangle$ and $|s_{-1}\rangle$. This kind of entanglement is, in principle, useful for phase encoding schemes of MA-MDI-QKD \cite{IEEE1}. In \cite{NVexp2}, the reported success probability or this entangling procedure is on the order of $10^{-4}$, which is higher than that of \cite{NVexp1}. The challenge here is that, in the phase encoding scheme, we need two QMs per users. If we entangle these two QMs using the setup in \cite{NVexp2}, it is possible to have double excitations, that is, to end up with two NV centers in their $|s_0\rangle$ states. In \cite{NVexp2}, the authors propose to flip the states of the NV centers and do the entangling procedure again. The desired entangled state would again emit a single photon, whereas the double-excited term, after flipping, would generate none. This way, we can basically purify our state to achieve high-fidelity entanglement. The price to pay is that by using the entangling procedure twice, the efficiency of the whole process would scale as the square of the single-stage entangling efficiency, which will be on the same order of magnitude as the scheme in \cite{NVexp1}. As for reading, while it is possible to apply the entangling procedure to double encode a photon with memories, it is challenging to perform an $X$-basis measurement on two separate NV centers. So, again, we end up with a scheme, which is neither efficient, nor a proper readout mechanism can be devised for it.

While the previous two schemes struggle with achieving high entangling efficiencies, partly because of their imperfect collection of the released photon, in \cite{Wrachtrup}, the authors report on a rather efficient, at around 20\%, {\em heralded} transfer of a single photon to the nuclear spin of an NV center. This scheme can, in principle, be employed in the setups of Figs.~\ref{fig:mem_assisted_setups}(a) and (c). The challenge with the setup in Fig.~\ref{fig:mem_assisted_setups}(a) is the rather long preparation time, on the order of 100~$\mu$s, in this scheme, which restricts the repetition rate of the protocol to below 10~kHz. If we switch to the setup of Fig.~\ref{fig:mem_assisted_setups}(c), which allows for delayed writing, the challenge would be in finding a high-rate EPR source by which the NV centers can be driven. The latter is non-trivial because cavity enhancement is often required for narrow-band high-rate sources. Finally, similar to the other two schemes, it is not at all obvious, how one can either read or double-encode the QMs with photons for the middle BSM operation.  

\subsubsection{Toy models for MA-MDI-QKD with non-cavity NV centers}
\label{toymodel}
While, in the lack of a proper readout scheme, we are not in a position to devise a full MA-MDI-QKD scheme for any of the above setups, we can still overestimate their performance by introducing a toy model that captures their key features. This model will not necessarily include all possible imperfections in such hypothetical setups, but, by that token, the key rates obtained from this model will provide us with an upper bound on the rate one can possibly achieve from such non-cavity setups. If this upper bound is still below the rate that no-QM systems offer, we can conclude that, in the context of MA-MDI-QKD, without cavity enhancement, our existing technology for NV centers is not capable of beating the no-QM systems. We discuss this further in Sec.~\ref{sec:results}.

In our toy model for the schemes in \cite{NVexp1} and \cite{NVexp2}, we assume that the entangled state in Eq.~\eqref{eq:NV_center1}
is always generated but because of the imperfect collection efficiency, the generated photon is directed to the side BSM with probability $p_{c}$. This way we ignore some of the other non-idealities that may bring the fidelity down. The resulting density matrix for the NV center (NV) and the collected photon (P) is then given by
\begin{equation}
\rho_{\mathrm{NV-P}}=p_{c}|\Psi_{1}\left\rangle \left\langle \Psi_{1}|+(1-p_{c})|{0}\left\rangle _{\mathrm{PP}}\hspace{-1mm}\left\langle {0}|\otimes I_{\mathrm{NV}}\right.\right.\right.\right. ,\label{eq:d_mat_first_scheme}
\end{equation}
where $I_{\mathrm{NV}}=(|0\rangle_{\rm NV} \langle 0|+|1\rangle_{\rm NV} \langle 1|)/2$
and $|0\left\rangle _{\mathrm{P}}\right.$ represents the vacuum state for the collected photonic mode. Note that for the scheme of \cite{NVexp2}, we need two NV centers on each side, namely, NV1 and NV2, for which $|0\rangle_{\rm NV} = |s_0\rangle_{\rm NV1} |s_{-1}\rangle_{\rm NV2}$ and $|1\rangle_{\rm NV} = |s_{-1}\rangle_{\rm NV1} |s_0\rangle_{\rm NV2}$. We use the term NC1 to refer to MA-MDI-QKD schemes that rely on the above entangling procedure.  

For the scheme proposed in \cite{Wrachtrup}, for which the setup in Fig.~\ref{fig:mem_assisted_setups}(c) is the most appropriate, we assume that an ideal EPR source with a matching bandwidth to the NV center is used. In Sec.~\ref{sec:results}, we use the specifications of single-photon sources that rely on NV centers to overestimate the rate parameters of such an EPR source. We use the term NC2 to refer to such an MA-MDI-QKD scheme.

Despite the fact that we are not aware of any readout mechanism by which the middle BSM can be done, for all three setups, we assume that one may come up with a reading protocol, with an efficiency $\eta_r$, by which the states of the QMs can be transferred to the photons. In that case, $\eta_r$ cannot be higher than the collection efficiency from a non-cavity memory.

Next, we consider MA-MDI-QKD with cavity-enhanced NV centers. 

\subsection{MDI-QKD with cavity-enhanced NV centers}

In this section, we propose an MA-MDI-QKD scheme that relies on cavity-enhanced NV centers as QMs. The key building block is an NV center whose internal state affects the effective reflectivity of the embedding cavity \cite{Bill_paper}. This idea of conditional reflection was first proposed in {\cite{duan_kible}} for a trapped atom system. Figure \ref{fig:scheme1_en_lev}(c) shows the relevant energy level structure for the NV center. Here, the resonant frequencies for $|s_{0}\left\rangle \right.\rightarrow|E_{x}\left\rangle\right.$ and {$|s_{+1}\left\rangle \right.\rightarrow|M_{5}\left\rangle \right.$} transitions are different and are, respectively, denoted by $\omega_{0}$ and $\omega_{1}=\omega_{0}+\delta$ . In \cite{Bill_paper}, authors assume that the NV center is embedded in a double-sided cavity with resonance frequency $\omega_{C}$ and reflectivities $r_{1}$ and $r_{2}$ for, respectively, input and output mirrors, and that the cavity is on resonance with $|s_{0}\left\rangle \right.\rightarrow|E_{x}\left\rangle\right.$ transition. They use this feature to perform conditional operations depending on the state of the NV center. We use the same idea but in the special case of a one-sided cavity. 

In general, the interaction of a single photon with the composite NV center-cavity system can be modeled by calculating the reflection amplitude, $A_{r}$, off the cavity, and transmission amplitude, $A_{t}$, through the cavity. For an incoming photon with frequency $\omega_{P}$, a cavity with resonance frequency $\omega_{C}$, and a two-level system embedded into the cavity with resonance frequency $\omega_{i}$, these amplitudes are given by \cite{Bill_paper}
\begin{eqnarray}
&A_{r}=1-\frac{1-A}{(1-i\Delta_{C})+2C/(1-i\Delta_{E})},& \nonumber\\
&A_{t}=\frac{\sqrt{1-A^{2}}}{(1-i\Delta_{C})+2C/(1-i\Delta_{E})},&
\label{eq:in_out_rel}
\end{eqnarray}
where $\Delta_{C}=\left(\omega_{P}-\omega_{C}\right)/\kappa,$ with $\kappa$ being the cavity decay rate, $\Delta_{E}=\left(\omega_{P}-\omega_{i}\right)/\gamma,$ with $\gamma$ being the spontaneous decay rate, and $C=\frac{g^{2}}{\kappa\gamma}$ is the cooperativity with $g$ being the coupling rate between the two-level system and the cavity mode. In Eq.~\eqref{eq:in_out_rel}, $A=\frac{r_{1}-r_{2}}{1-r_{1}r_{2}}$ is the amplitude of the reflected light for an empty cavity on resonance. In the following, by considering special cases for the above parameters, we come up with a new entangling technique for NV-center-based MA-MDI-QKD. 

\subsubsection{Our proposed polarization encoding scheme}
\label{Sec:ourscheme}

\begin{figure}
\begin{centering}
\includegraphics[scale=0.7]{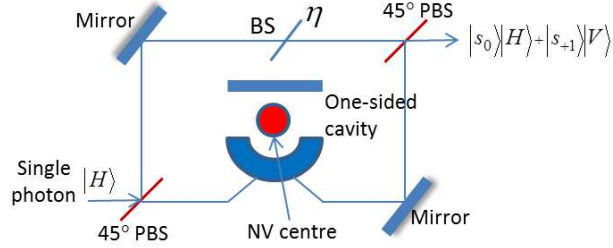}
\par\end{centering}
\protect\caption{\label{fig:Entangling-a-polarized}The double-encoding module in our proposed scheme. It entangles a polarized photon with an NV center in a cavity. This module will be used for initialization, encoding, and the final readout operations.}
\end{figure}

In this section, we describe our proposed MA-MDI-QKD scheme that relies on NV centers embedded into small-volume cavities. The key enabling idea is to treat the cavity-NV center as a conditional reflection module. To that end, suppose NV centers are embedded into one-sided cavities, i.e., $r_{2}=1$, and the cavity mode is on or near resonance with the incoming photon, both on or near resonant with the $|s_{0}\left\rangle \right.\rightarrow|E_{x}\left\rangle\right.$ transition, i.e., $\omega_{C}\sim \omega_{P}=\omega_{0}$. Under this condition, Eq.~\eqref{eq:in_out_rel} reduces to
{
\begin{equation}
A_{r}=1-\frac{2}{(1-i\Delta_{C})+2C/(1-i\Delta_{E})}, \;\;\;\;\;\;\;\;\;\;\;\;
A_{t}=0.
\label{eq:ref_trans_coeff}
\end{equation}
}
\noindent In Eq.~\eqref{eq:ref_trans_coeff}, if the NV center is in state $|s_{0}\left\rangle \right.$, we have $\Delta_{E}=0$,
which, for $C\gg 1$, would result in $A_{r}\sim1$, i.e., the photon will be reflected off the cavity as if it has hit a mirror. This is because, in this case, the incoming pulse is detuned from the frequency of the dressed cavity mode \cite{duan_kible}. When the NV center is in state $|s_{+1}\left\rangle \right.$, however, $\Delta_{E}=-\delta/\gamma$. But, assuming that $\delta\gg\gamma C,$ we end up with $A_{r}\sim-1$. This implies that in both cases the photon will get reflected but it will acquire different phase shifts depending on the state of the NV center. Obviously, for finite values of $C$ and $\delta$, we may deviate from this ideal scenario. We will study the implications of such realistic cases later. For now, let us carry on with the ideal picture to describe our key entangling scheme.

{\em Double encoding:} The key building block in our scheme, which will be used in all three stages of initialization, loading, and reading the memory, is the double-encoding module in Fig.~\ref{fig:Entangling-a-polarized}. This module uses the above-mentioned conditional phase gate to entangle the polarization of a single photon with the electron spin of an NV center. It ideally works as follows. Suppose the NV center has been initialized to the state $|\Psi_{in}\left\rangle =\left(|s_{0}\left\rangle \right.+|s_{+1}\left\rangle \right.\right)/\sqrt{2}\right.$; the initialization procedure will be explained later in this section. We then generate an $H$-polarized single photon with frequency $\omega_0$ and send it through a $+45^{\circ}$ polarizing beam splitter (PBS). We can generate such a single photon by driving the $|s_{0}\left\rangle \right.\rightarrow|E_{x}\left\rangle\right.$ transition in another cavity-NV-center pair. In Fig.~\ref{fig:Entangling-a-polarized}, the $+45^{\circ}$-polarized component of this single photon interacts with the NV center, resulting in the joint state {$|D\left\rangle_{s}\right.\left(|s_{0}\left\rangle \right.-|s_{+1}\left\rangle \right.\right)/\sqrt{2}$, where $|D\rangle =\frac{1}{\sqrt{2}}\left(|H\rangle+|V\rangle \right) $.} The photonic modes $r$ and $s$ are then recombined at a second {$+45^{\circ}$ PBS, which will ideally result in the following output state
\begin{equation}
|\Psi_{2}\left\rangle \right.=\frac{1}{\sqrt{2}}\left(|H\left\rangle \right.|s_{0}\left\rangle \right.+|V\left\rangle \right.|s_{+1}\left\rangle \right.\right).\label{eq:double-enc}
\end{equation}
Here, the interaction time, $\tau_{\mathrm{int}}$, corresponding to the above double-encoding procedure is expected to be about 10~ns.}

{In deriving Eq.~\eqref{eq:double-enc}, we have made the assumption that the reflection coefficient in Eq.~\eqref{eq:in_out_rel}, in the two cases of $|s_0\rangle$ and $|s_{+1}\rangle$ states, has the same magnitude of 1. For finite values of $C$, however, the two coefficients may not take their ideal values, and this would result in a deviation from the ideal entangled state in Eq.~\eqref{eq:double-enc}. For instance, at $C=50$, $\Delta_C=-1$, and $\Delta_E=-100$, we have $A_{r}(|s_{0}\rangle) \sim A_{r}(|s_{+1}\rangle) \sim 0.98$. This will cause an imbalance between the two legs of the interferometer in Fig.~\ref{fig:Entangling-a-polarized}. We can fix this by adding a beam splitter with transmissivity $\eta$ in the $r$ branch. The value of $\eta$ will be chosen accordingly to account for different sources of loss in the $s$ branch. In this case, the generated state by our double-encoder will become
\begin{equation}
\rho_{\mathrm{NV-P}}=\eta|\Psi_{2}\left\rangle \right.\left\langle \right.\Psi_{2}|+(1-\eta)|\mathbf{0}\left\rangle \negthinspace\negthinspace\right.{}_{\mathrm{PP}}\hspace{-1mm}\left\langle \right.\mathbf{0}|\otimes I_{\mathrm{NV}}^{'},\label{eq:den_mat_sec_scheme}
\end{equation}
where $I_{\mathrm{NV}}^{'}=(|s_{0}\rangle \langle s_{0}|+|s_{+1}\rangle \langle s_{+1}|)/2$. This is similar to Eq. \eqref{eq:d_mat_first_scheme}, with the difference that now $\eta$ can be several orders of magnitude larger than $p_{c}$. With the above state, we expect that the user's state will be properly teleported to the QM in majority of cases where the side-BSM has been successful, i.e., two detectors have clicked. The vacuum state in Eq.~\eqref{eq:den_mat_sec_scheme} ideally should not result in a successful side-BSM. But, with a rate proportional to the detector's dark count rate, we may still get erroneous side-BSM results that may induce errors in the end. Although small, we consider this effect in our key rate calculations in Sec. IV. 

There are other practical points to consider with regard to the entangling scheme of Fig.~\ref{fig:Entangling-a-polarized}. In Eq.~\eqref{eq:den_mat_sec_scheme}, we assume that the reflection coefficients for $|s_0\rangle$ and $|s_{+1}\rangle$ have the same magnitude, although not necessarily one. Depending on the actual parameter values that the implementation of our cavity system offers, this may not always be possible. For instance, at $C=50$, $\Delta_C=-0.33$, and $\Delta_E=-300$, $A_{r}(|s_{0}\rangle)=0.98$, whereas $A_{r}(|s_{+1}\rangle) \sim 1$. The imbalance would be higher at lower values of $C$, which represent a more practical regime of operation. In Sec.~\ref{sec:results}, we study how the key rate drops as a result of this imbalance and find out the minimum value of $C$ at which our system still offers some advantage. Another issue with less then unity reflection coefficients is the possibility of the photon entering the cavity, being absorbed by the NV center and then being non-radiatively emitted . For the NV center in $|s_0\rangle$, the chance of this happening is about 1\% of the cases that the photon is not directly reflected off the cavity. The latter will happen with probability $1-\eta_{r0}$, where $\eta_{r0} = |A_{r}(|s_{0}\rangle)|^2$. For instance, for the numerical example above, $1-\eta_{r0}=0.04$, and therefore the chance of non-radiative emission is only 0.04\%. While the rate at which this may occur is rather low, once it happens, the NV center may stay in certain undesired metastable states for 250-500~ns, during which we cannot initialize the memory in the desired state. This will result in a certain deadtime, $\tau_{\rm dead}$, for our scheme during which we cannot teleport the user's photon to its respective QM. In our key rate analysis, we account for this effect by a correction factor that modifies the probability of loading in our setup.


{\em Initialization:} Before performing the double encoding operation above, at the beginning of each round, we need to first initialize the NV center in state $|\Psi_{in}\rangle$. This can be done by the double-encoding module of Fig.~\ref{fig:Entangling-a-polarized}. In every round, we drive the NV-center-cavity module by an {$H$-polarized} single photon, and measure the polarization of the output photon in $|H\rangle$ and $|V\rangle$ basis. If we get a click, that would correspondingly project the NV center to $|s_0\rangle$ or $|s_{+1}\rangle$ states. We can then apply the relevant rotation to initialize the NV center in $|\Psi_{in}\rangle$. Each round of the above procedure includes the double encoding operation and then a rotation. This altogether roughly takes 15~ns \cite{Bill_paper} and corresponds to the initialization time, $\tau_{\mathrm{init}}$, in our protocol. 

In the above procedure, if we get no click, then our initialization has failed. If this happens for several consecutive rounds, that would indicate that the memory is in a deadtime period. During the deadtime, the NV center is in certain metastable states, which can decay to any of $|s_{0}\rangle$ and $|s_{\pm 1}\rangle$ states. Given that $|s_{-1}\rangle$ is not in the desired manifold of states that we need, during the deadtime, we swap states $|s_0\rangle$ and $|s_{-1}\rangle$ in every initialization round to avoid the possibility of staying in $|s_{-1}\rangle$ for ever.



{\em Readout:} We use the double-encoding technique to read out the memories and perform the middle BSM. This can be done by the module of Fig.~\ref{fig:Entangling-a-polarized}. This would map the QM state $|s_{0}\left\rangle \right.$ to $|s_{0}\left\rangle |H\left\rangle ,\right.\right.$ $|s_{+1}\left\rangle \right.$ to $|s_{+1}\left\rangle |V\left\rangle ,\right.\right.$ and $|s_{0}\left\rangle \right.\pm|s_{+1}\left\rangle \right.$ to $|s_{0}\left\rangle |H\left\rangle \pm|s_{+1}\left\rangle |V\left\rangle .\right.\right.\right.\right.$ Charlie will also need to do an $X$-basis measurement in the $|s_{0}\left\rangle \right.\pm|s_{+1}\left\rangle \right.$ basis, on the NV centers and will send its results to the end users. The time needed for the readout operation is estimated to be around 25~ns, which includes the double-encoding time and the time needed for the $X$ measurement basis. The latter involves a $\pm\pi/2$ rotation followed by a $Z$-basis measurement on the electron spin state of the NV center.

There are several requirements for the above setup to work properly. First, we assume that strong coupling between the NV center and a microcavity can be established. This has not yet been demonstrated in the laboratory, but experimental efforts are underway. In our work, we estimate how strong this coupling should be. We show that, even with moderate coupling, our system can offer some advantages. Second, this setup requires nearly on-demand single-photon sources (SPSs) at the middle station. This is not, however, an additional requirement as once cavity embedded NV centers are fabricated, one can use them to generate single photons with matching bandwidths to our transitions of interest. In our rate analysis, we assume that the employed SPSs are probabilistic, but they generate true single photons. The latter assumption is crucial as, otherwise, the multi-photon errors at the middle station can be detrimental to the key objective behind MA-MDI-QKD \cite{IEEE1}. Finally, we need to maintain polarization across the channel, which can be challenging over long distances. This condition can be alleviated by using an equivalent phase-encoding scheme \cite{Bit_ass}.

In the next section we analytically calculate the secret key rate of our proposed scheme and we compare it with that of existing no-memory QKD schemes, the non-cavity NV centers, and several other memory candidates.

\section{Key rate analysis}
\label{sec:results}


In this section the secret key generation rate of the proposed setups in Sec.~\ref{Sec:withNV} is obtained under the normal operation condition when no eavesdropper is present. We assume that single-photon sources are used at the users' ends. This is not an essential assumption; it just provides a convenient approach to compare memory-assisted schemes with their no-QM counterparts. In practice, one can use decoy-state techniques, for which similar margins of improvement over decoy-state no-QM systems are expected. In \cite{panayi}, the total secret key generation rate, using the efficient QKD protocol when ideal single photon sources are used by the users and the $Z$ basis is more often used than the $X$ basis, is lower bounded by the following expression
\begin{equation}
R_{\mathrm{QM}}=\frac{R_{S}}{N_{L}\left(P_{A},P_{B}\right)+N_{r}}Y_{11}^{QM}(1-h(e_{11;X}^{\mathrm{QM}})-fh(e_{11;Z}^{\mathrm{QM}})),\label{eq:key_rate}
\end{equation}
where $P_{A}$ and $P_{B}$ represent the probability of a successful side-BSM on, respectively, Alice and Bob's side; $Y_{11}^{QM}$ is the probability that the middle BSM is successful assuming that both memories are loaded (in the Z basis); $e_{11;X}^{\mathrm{QM}}$ and $e_{11;Z}^{\mathrm{QM}}$ are, respectively, the quantum bit error rate (QBER) between Alice and Bob in the $X$ and $Z$ basis when single photons are sent by the users; $f$ is the inefficiency of error correction; $h(q)=-q\log_{2}q-(1-q)\log_{2}(1-q)$ is the binary entropy function; $R_{S}=1/T$ is the repetition rate; $N_{L}$ is the average number of trials to load both memories, which is approximated by $3/(2P_{A})$ when $P_{A}=P_{B}\ll1$; and, $N_{r}=\left[\frac{\tau_{w}+\tau_{r}}{T}\right]-1$ is the number of rounds that we lose from the time that both memories are loaded until we learn the result of the final BSM operation. In \cite{panayi}, $\tau_w$ and $\tau_r$, respectively refer to writing and reading times. In our work, we provide a more detailed description of these parameters, specific to protocols used, as follows:
\begin{itemize}
\item In all schemes that use the setup of Fig.~\ref{fig:mem_assisted_setups}(b), the entire protocol can be run at a period given by $\tau_{w} =\tau_{\mathrm{init}} + \tau_{\mathrm{int}} +\tau_{\rm M} = T$, where $\tau_{\rm M}$ is the verification time at the side-BSMs, which is expected to be around 1~ns, hence negligible as compared to the other two terms. In this scheme, $\tau_r = \tau_{\mathrm{int}} + \tau_M +\tau_{\rm PM}$, where $\tau_{\rm PM}$ is the time required for any post-measurement operation, such as the $X$-basis measurement in the double encoding technique.
\item In all schemes that use the setup of Fig.~\ref{fig:mem_assisted_setups}(c), $\tau_w = T$, where the latter is determined by the rate at which the slower of EPR source and the user's source can be driven. For the delayed writing scheme, $\tau_r = 2\tau_M + \tau_{\rm int} + \tau_{\rm init}$. The reason for this is as follows. In the scheme of Fig.~\ref{fig:mem_assisted_setups}(b), we write onto the memories in every round. That is why we have to initialize the memory before the next photon arrives. In Fig.~\ref{fig:mem_assisted_setups}(c), we only write into the memory when we have a successful side-BSM. In this case, the initialization can be done once the memory is read for the middle BSM. That is why $\tau_{\rm init}$ is part of $\tau_r$ in this scenario.
\end{itemize}


We have used the machinery developed in \cite{panayi} and \cite{IEEE1} to find the key parameters in Eq.~\eqref{eq:key_rate} the details of which appear in Appendix A. The derivations are cumbersome and have mostly been done by the symbolic software Maple. In short, for each scheme, we first obtain the state of the QMs once the user's state is loaded to them. Our calculation includes all loss elements, dark count, and all the nonidealities we modeled in the entangling procedures in Sec. III. At this stage, we also find $P_{A}$ and $P_{B}$. In order to do so, we first find these parameters assuming that the deadtime is zero. Denote the loading probabilities in this latter case by $P_{A0}$ and $P_{B0}$. Accounting for the deadtime issue, we then get $P_{A}= P_{A0}(1-N_{\rm dead}p_{\rm dead})$ and $P_{B}=P_{B0}(1-N_{\rm dead}p_{\rm dead})$, where $N_{\rm dead}=\tau_{\rm dead}/T$ is the number of rounds lost to the deadtime, and $p_{\rm dead} = 0.01(1-\eta_{r0})$. In the last expression, 0.01 is the probability of transition to metastable states from $|s_0\rangle$; the chance of being in $|s_0\rangle$ is assumed to be 1/2; and we have accounted for the possibility of entering deadtime either at the initialization stage or the double-encoding stage. We have neglected the deadtime cases arising from the final BSM procedure, as the number of times that this happens is considerably lower than the former two processes, which are used in every round. We then model memory decoherence using a depolarizing channel, see Eq.~\eqref{eq:dec_d_mat}, with a time constant $T_{2}$.  {This is perhaps a conservative assumption for NV centers, but it agrees with the analysis reported in \cite{NV2}}. We assume that the amplitude decay time, $T_1$, is sufficiently large in all schemes considered in this paper. We then model the final BSM on the decohered states of the QMs, taking into account the statistics of loading. As a result, we can calculate the remaining terms in Eq.~\eqref{eq:key_rate}, i.e., $Y_{11}^{QM},$ $e_{11;X}^{\mathrm{QM}}$, and $e_{11;Z}^{\mathrm{QM}}$.

\subsection{Numerical Results}

\begin{table}
\footnotesize
\begin{centering}
\textcolor{black}{}%
\begin{tabular}{|c|c|c|c|c|c|c|}
\hline 
 & \textcolor{black}{Our NV} & \textcolor{black}{quantum} & \textcolor{black}{trapped} & \textcolor{black}{trapped} & NC1 & NC2\tabularnewline
  & \textcolor{black}{scheme} & \textcolor{black}{dots} & \textcolor{black}{atoms} & \textcolor{black}{ions} &  & \tabularnewline
\hline 
\textcolor{black}{Entangling efficiency, $\eta$, $p_c$} & \textcolor{black}{$0.9$} & \textcolor{black}{$0.9$} & \textcolor{black}{NA} & \textcolor{black}{NA} & $10^{-3}$ & NA\tabularnewline
\hline 
\textcolor{black}{Writing efficiency, $\eta_{{w}}$} & NA & NA & 0.39 & $1$ & NA & 0.2\tabularnewline
\hline 
\textcolor{black}{Reading efficiency, $\eta_{{r}}$} & NA & NA & 0.69 & $1$ & 0.2 & 0.2\tabularnewline
\hline 
\textcolor{black}{Up-conversion efficiency} & \textcolor{black}{0.68} & \textcolor{black}{1} & \textcolor{black}{1} & \textcolor{black}{0.68} & 1 & 1\tabularnewline
\hline 
\textcolor{black}{Coherence time, $T_{2}$} & \textcolor{black}{10-100 ms} & \textcolor{black}{1 $\mu$s} & \textcolor{black}{1 $\mathrm{s}$} & \textcolor{black}{50 s} & 10 ms & 10 s\tabularnewline
\hline 
Repetition rate, $R_{S}$ & 40 MHz & 100 MHz & 10 MHz & 10 MHz & 7 MHz & 200 kHz\tabularnewline
\hline 
Interaction time, $\tau_{\rm init}$ & 10 ns & $5$ ns & 10 $\mu$s & 10 $\mu$s & 10 ns & 10 ns\tabularnewline
\hline 
Initialization time, $\tau_{\rm int}$ & 14 ns & 5 ns & 144.7 $\mu\mathrm{s}$ & 120 ms & 7 $\mu\mathrm{s}$ & 100 $\mu\mathrm{s}$\tabularnewline
\hline 
Verification time, $\tau_{\rm M}$ & 1 ns & 1 ns & 1 ns & 1 ns & 1 ns & 1 ns\tabularnewline
\hline 
Post-meas. time, $\tau_{\rm PM}$& 14 ns & 0  & NA & 0 & NA & NA\tabularnewline
\hline
\end{tabular}
\par\end{centering}

\protect\caption{\label{tab:Nominal-values-used}Nominal values used in our numerical
results for different platforms and setups. In all setups, we assume a single-photon detector efficiency of 0.93 and a dark count rate of 1 cps \cite{dark_count}. The single-photon sources have an efficiency of 0.72 per trigger, and the attenuation length of the channel, $L_{\rm att}$ is 25~km. In the case of trapped-ions, trapped atoms, and quantum dots {we have used the nominal
values reported in \cite{TrappedIon}, \cite{Rempe}, and \cite{QDotSource}, respectively. If a parameter value has not been available, an appropriate estimate has been used. NA means not applicable. NC1 and NC2 refer to the non-cavity cases modeled in Sec.~\ref{toymodel}.}}
\end{table}

In this section, we compare the rate of our proposed cavity-based MA-MDI-QKD scheme with that of non-cavity models, NC1 and NC2, as well as a range of other single-excitation QMs, namely, quantum dots, trapped atoms and trapped ions. In all cases, we compare the rate with that of a no-QM MDI-QKD system driven at a 1~GHz rate. We also compare our system with the no-QM setup proposed in \cite{Azuma2015}, which relies on linear optics and quantum non-demolition (QND) measurement. The nominal values used in each case is summarized in Table~\ref{tab:Nominal-values-used}. In the case of quantum dots, we use the results reported in \cite{QDotSource} for spin-photon entanglement, to estimate the time parameters in a corresponding MA-MDI-QKD scheme as in Fig.~\ref{fig:mem_assisted_setups}(b). The entangling efficiency has, however, been boosted to what we assume in our cavity-based scheme with NV centers for fair comparison. For trapped atoms, we use the results reported in \cite{Rempe} to calculate the key rate of a corresponding MA-MDI-QKD scheme as in Fig.~\ref{fig:mem_assisted_setups}(c). Considering the rather long initialization time for trapped atoms, the rate for the EPR-based scheme with delayed writing will be higher than the double-encoding scheme of Fig.~\ref{fig:mem_assisted_setups}(b). The same holds for trapped-ion based MA-MDI-QKD, for which relevant parameters are taken from \cite{TrappedIon}. In the case of trapped atoms or ions, we have assumed that we can drive the system with a narrow-width EPR source at a 10 MHz rate. That would correspond to the spontaneous decay rate of a typical alkali atom used in such systems. Note that, in the case of trapped ions, the middle BSM can be performed deterministically. Once we have proper sources that can interact with our QMs, one should also consider the use of frequency converters to enable the interaction between a QM-driven photon and the telecom photon sent by the user. This could reduce the total efficiency of our BSM operations and is modeled as an additional source of loss. While, for each system, a proper up-converter needs to be designed and implemented, we estimate the efficiency of such up-converters by looking at similar examples in the literature \cite{upconversion,up_conv1}. We have used the ideal unity conversion efficiency for memory systems that have poorer performance than the no-QM setup, as it does not change the conclusion of our analysis.


\begin{figure}
\begin{centering}
\includegraphics[scale=0.3]{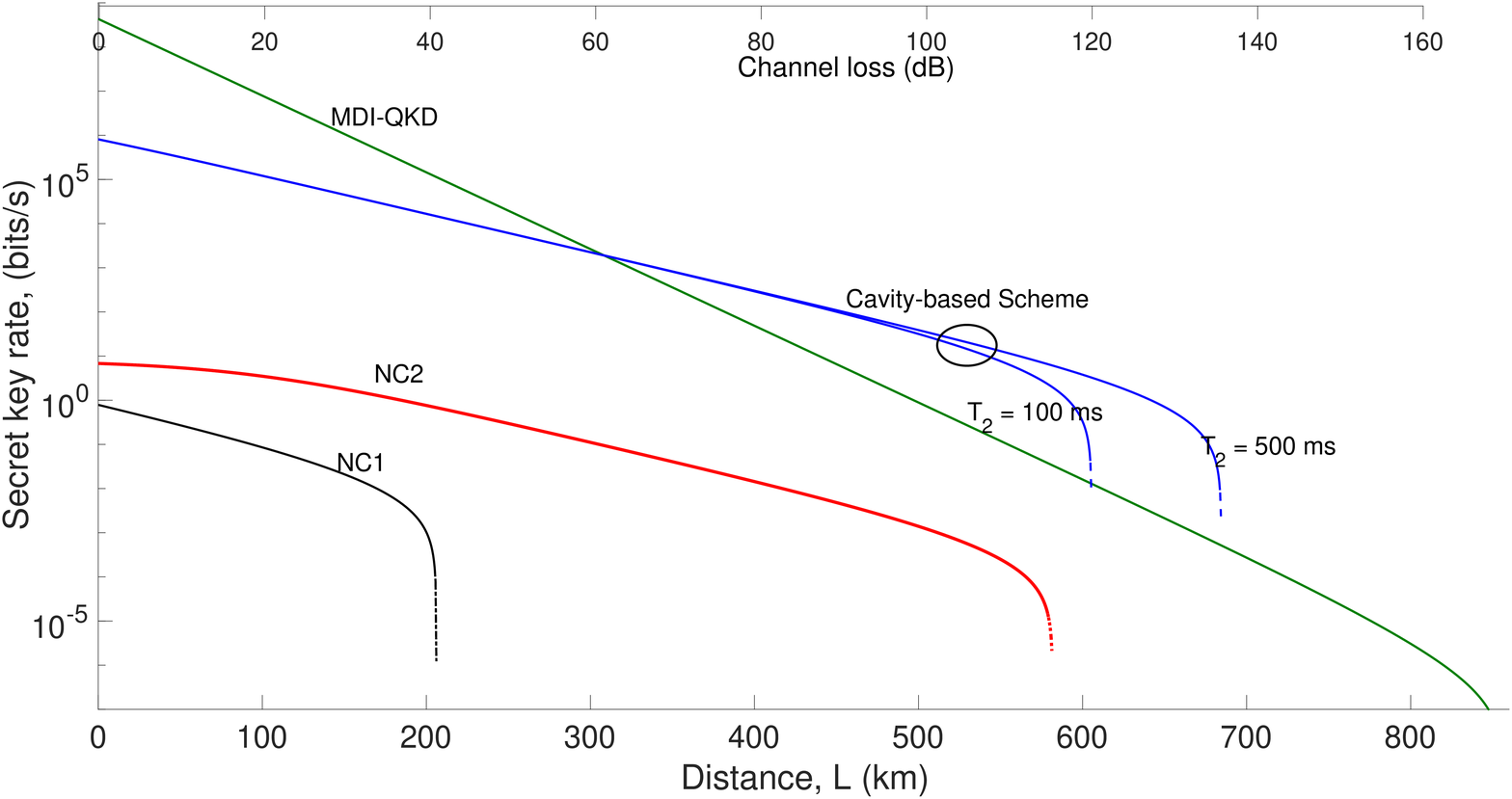}
\par\end{centering}
\protect\caption{\label{fig:Secret-key-generation_rate}Secret key generation rates versus distance for MA-MDI-QKD using non-cavity (NC1 and NC2) and cavity based NV centers, and its comparison with no-QM MDI-QKD driven at 1~GHz. Nominal values used are summarized in Table \ref{tab:Nominal-values-used}.}
\end{figure}

Figure \ref{fig:Secret-key-generation_rate} compares the secret key generation rate for all NV-center based schemes proposed in Sec.~ \ref{Sec:withNV}. In the case of NC1, which corresponds to what we can potentially get from the entangling schemes in \cite{NVexp1} and \cite{NVexp2}, we have assumed an entangling efficiency of $p_c=10^{-3}$ and a reading efficiency of 0.2, which are both optimistic assumptions for a non-cavity system. Despite of these generous parameter values, the key rate for the NC1 model is the worst of all systems  considered in Fig.~\ref{fig:Secret-key-generation_rate} and will not cross the no-QM curve. This is partly because of the low entangling efficiency and partly the low repetition rate resulting from the microsecond-long initialization. In the case of NC2, which relies on an ideal EPR source, we assume a 200 kHz repetition rate. This corresponds to the best rates reported for single-photon sources that rely on NV centers in nanowires \cite{Diamond_nanowire}. Because of using nuclear spins, the coherence time is much longer at 10 s. The other parameters are taken from \cite{Wrachtrup}. It can be seen that the NC2 curve cannot surpass the no-QM curve either. Note that the toy models NC1 and NC2 have already neglected many possible sources of error in the system, despite of which neither can outperform the no-QM system. We can then conclude that without using small-volume cavities, it may not be possible to outperform existing no-QM systems by using NV centers in the MA-MDI-QKD setups.

The situation above would change if we do have cavity-enhanced NV centers as we described in Sec.~\ref{Sec:ourscheme}. In Fig.~\ref{fig:Secret-key-generation_rate}, we have plotted the key rate for our proposed scheme at two different values of coherence time. At $T_2 = 100$~ms, which is the typical coherence time of electron spins \cite{T2_2}, we outperform the no-QM system by nearly one order of magnitude at distances around 400 km. We can extend the window over which our NV-center based scheme outperforms MDI-QKD if we increase the coherence time by one order of magnitude. {This is in principle possible, if one uses spin echo like techniques \cite{Toyofumi2012,Bar-Gill2013}, or transfer the electron spin to nuclear ones \cite{Yang2016}.} This implies that the cavity-based NV centers have the potential of beating no-QM systems over a distance range of interest. Note that for our scheme, we have used an entangling efficiency of $\eta = 0.9$ for our double-encoding module in Fig.~\ref{fig:Entangling-a-polarized}. The assumption here is that the two legs of the double encoder in Fig.~\ref{fig:Entangling-a-polarized} are balanced. With a cooperativity on the order of 50, we expect a reflectivity coefficient around 98\%, corresponding to $\eta = 0.96$. The 90\% efficiency will then account for other possible sources of loss in the double encoder as well. We have also accounted for the possible deadtime caused by overstaying in metastable states of the NV center in our loading parameters. We have assumed that $\tau_{\rm dead}=500$~ns, corresponding to $N_{\rm dead}=20$ rounds of our protocol. The same effects have been accounted for during the initialization of the QMs. The corresponding time parameters in our scheme are taken from the results reported in \cite{Bill_paper,rotational_errors}. 

\begin{figure}
\begin{centering}
\includegraphics[scale=0.8]{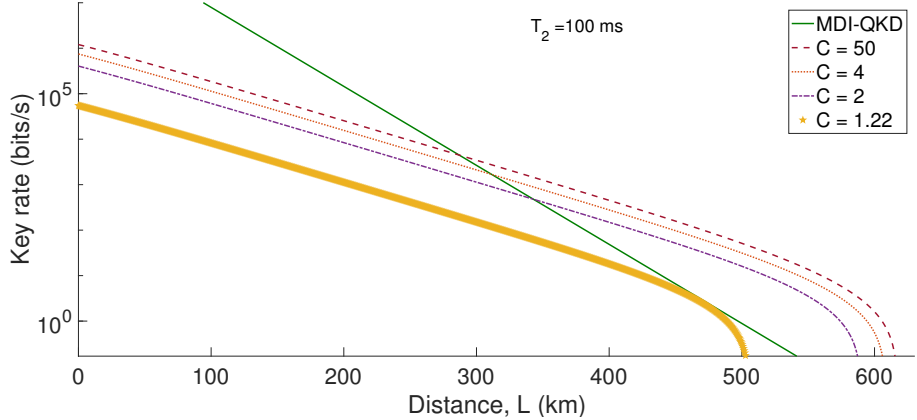}
\par\end{centering}
\protect\caption{\label{fig:diff_C} The total secret key generation
rate of MA-MDI-QKD for cavity-based NV centers for different values of cooperativity. In all curves, $\eta =1$, $\Delta_E = -300$, and $\Delta_C = 2 C \Delta_E/(1+\Delta_E^2)$. For the range of values considered for $C$, the choice of value for $\Delta_C$ would result in mainly real values for $A_r(|s_0\rangle)$ and a nearly unity value for $A_r(|s_1\rangle)$.}
\end{figure}

While it is promising that we can beat no-QM setups using cavity-based NV centers in their strong coupling regime, it is important for experimentalists to know how strong this coupling should be. For that matter it is necessary that we calculate the key rate for more realistic parameter values. Figure~\ref{fig:diff_C} provides an answer to this. In this figure, we use $\Delta_E = -300$, corresponding to a typical NV center, and, for each value of $C$, we tune $\Delta_C$ to give us real values for $A_r(|s_1\rangle)$. This setting results in an imbalanced setup in which $A_{r1}=A_r(|s_1\rangle) \approx 1$, and $A_{r0}=A_r(|s_0\rangle)<1$. In all curves in Fig.~\ref{fig:diff_C}, we then consider a simplified setup in which $\eta = 1$. In principle, one can optimize $\eta$ to get even higher rates. The result is quite promising: with even low values of $C$ on the order of 1 we can still beat a typical no-QM QKD system. By writing the full state of the system, we can show that when $A_{r0}=A_{r1}$, the QBER roughly scales with $(1-A_{r0})^2$, whereas in the imbalanced case of $A_{r0}<A_{r1}=1$, there are terms that scale with $(1-A_{r0})$. Once $A_{r0}$ goes down, these terms bring the total key rate down to the point that we can no longer outperform a conventional QKD system.

Figure \ref{fig:Comp_with_no_mem} compares the total secret key generation rate for various candidates for QMs. These cases include trapped atoms in optical cavities, trapped ions, and quantum dots and NV centers embedded into small volume cavities. These examples would represent the major memory candidates with single-excitation features. For each memory we use the setup that offers the highest key rate, although improvements may still be possible if one further investigates the specific features of each QM. In terms of initialization times, quantum dots are the fastest of all, but their coherence time is often too low, which results in their fast decoherence before they get to outperform the no-QM system. The problem with low coherence times can potentially be alleviated if one uses the multiplexing idea in multiple-memory scenarios \cite{Kuzmich_QMmux, mult_mem}. That would, however, add to the complexity of the implementation. For the slower trapped-atom and ion QMs, we need to have a proper EPR source to have a chance at beating no-QM systems. For an EPR source driven at 10 MHz, trapped-atom QMs would also fall short of taking over the no-QM system if their coherence time is limited to 1 s. Trapped ions, with typically much longer coherence times, have the potential to beat no-QM systems, but that only happens at rather long distances and very low rates on the order of 1 b/s. The latter is because of their 100-ms-long time parameters, which, in the absence of any inefficiencies, would limit their key rate to 10 b/s. Note that we have assumed ideal reading and writing efficiencies for trapped-ion QMs. Among all the QM options we have considered, the NV centers seem to be the only one that can offer some advantage over no-QM systems in a practical regime of interest. 

\begin{figure}
\begin{centering}
\includegraphics[scale=0.3]{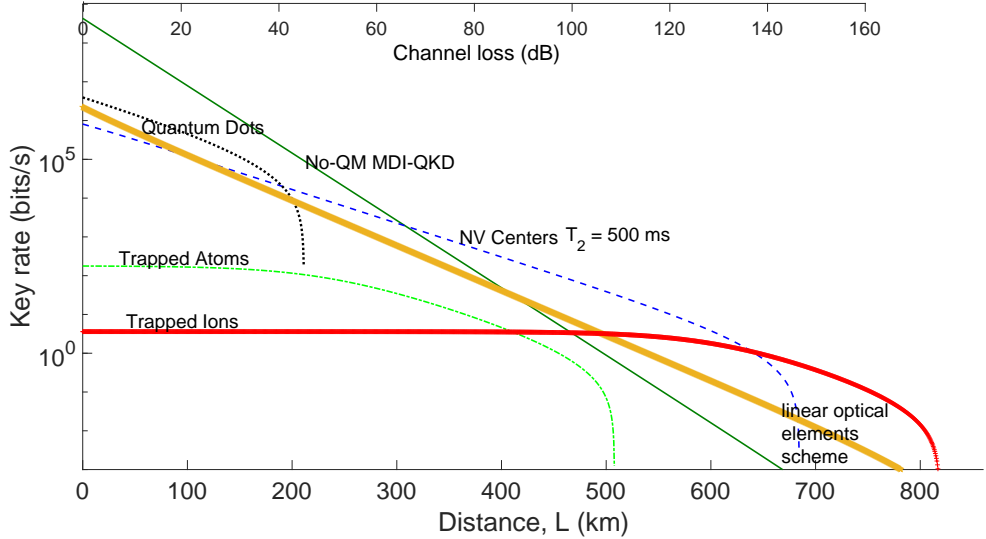}
\par\end{centering}
\protect\caption{\label{fig:Comp_with_no_mem}Comparison of the total secret key generation
rate versus distance for the MA-MDI-QKD schemes relying on cavity-based NV centers and quantum dots versus trapped atoms and ions. We also compare our performance with that of the linear optical elements scheme proposed in \cite{Azuma2015}. The relevant system parameters are given in the text and in Table~\ref{tab:Nominal-values-used}.}
\end{figure}

Finally, in Fig.~\ref{fig:Comp_with_no_mem}, we also compare the performance of our proposed NV-center based systems with the no-QM setup proposed in \cite{Azuma2015}. In \cite{Azuma2015}, authors propose to replace the QM modules in Fig.~\ref{fig:mem_assisted_setups}(a) with QND modules, and then run a large number of such systems in parallel. Using QND measurements, they can tell, which photons have survived the path loss, on which, using a fast optical switch, they perform a BSM. They then show that the normalized rate by the total number of systems used, $N$, scales the same as that of MA-MDI-QKD. In order to compare this system with the NV center one, we have to make some assumptions on how the former will be implemented. In our comparison, we assume that the QND operation is implemented using the teleportation idea in Fig.~\ref{fig:mem_assisted_setups}(c) that relies on an EPR source. The non-measured EPR photon will then be sent to a large switch, instead of the QM, to be used for the central BSM if the QND is successful. Another assumption we make is the inclusion of insertion loss in such fast, but single-photon level, optical switches. The typical problem with such switches is that they are often too lossy, with up to 3 dB loss for a $2\times2$ switch with nanosecond switching time. Here, we assume a switching time of 10~ns, hence a repetition rate of 100~MHz, with an equivalent insertion loss of 0.5~dB for a $2\times2$ switch. For an $N$-port switch, the total insertion loss would be given by $0.5 \log_2 (N)$~dB. We consider this loss factor in our calculation of the key rate. Finally, the ideal rate-versus-distance behavior occurs when $N$ is large. In our simulation, we have assumed $N = 1/P_A$. All put together, the curve labeled linear optical elements scheme shows the performance of the system proposed in \cite{Azuma2015}. Within the employed assumptions, the NV center based system performs better than that of the linear optical scheme. We should also bear in mind that for MA-MDI-QKD, we only need to implement and run one setup, whereas for the proposal in \cite{Azuma2015}, we need a large number of parallel systems. For instance, at $L= 400$~km, where the linear optical scheme starts outperforming the conventional no-QM systems, we need around 10,000 parallel systems, which makes the implementation of such systems challenging. This may suggest that, in short term, the MA-MDI-QKD has a better chance at improving the rate-versus-distance behavior than its rivals.

\section{Conclusions and discussions}
\label{Sec:Conc}
In this paper, we studied the suitability of NV centers in diamond as memories in MA-MDI-QKD systems. We considered several experimental setups, in all of which the NV center interacts with a free-space photon, versus theoretical proposals that rely on NV centers embedded into small-volume optical cavities. The key objective was to find a regime of operation that the MA-MDI-QKD system could outperform no-QM counterparts. It turned out that, even by making optimistic assumptions, the no-cavity systems were not able to beat the no-QM systems. With cavity enhancement, however, our proposed scheme could outperform the original MDI-QKD over roughly 300-500 km. Most importantly, the required cooperativity for such cavity coupling was shown to be on the order of one. In comparison with other single-excitation QMs, such as quantum dots, trapped atoms and trapped ions, cavity-based NV centers had the potential to be the most practical candidate for beating conventional QKD demonstrations.

Our analysis is based on certain assumptions on the capabilities that may only be available in the near future. This is not per say unacceptable, noting that we do not have, at the moment, a working family of QMs suitable for MA-MDI-QKD. But, like NV centers, each requires to become maturer in order to offer a practical advantage over no-QM systems. In the case of NV centers, our results show that embedding NV centers into microcavities is a must, given that the no-cavity setups we considered were not able to offer any advantages. While progress is being made by several groups worldwide, such a device is yet to be fabricated. Nevertheless, the required cooperativity values seem to be within reach of early demonstrations. Our proposed scheme also requires a near deterministic high-rate single-photon source matched to our NV center. Here, we are fortunate as the NV centers embedded in
the microcavity can also act as a single photon source, offering extremely small multi-photon rate as needed for MA-MDI-QKD. Note that with such single-photon sources one can devise alternative setups for ensemble-based QMs, which do not suffer from the multiple-excitation issue \cite{LoPiparo:CLEO16}.  Finally, for the state-dependent optical coupling required in our scheme, low temperature operation at around 4-8 K is required.

While MA-MDI-QKD is potentially capable of beating existing no-QM systems over a range of distances, for a no-limit trust-free long-distance QKD, one eventually needs to use quantum repeater structures \cite{briegel,IEEE2}. MA-MDI-QKD, nevertheless, provides an intermediary solution compatible with the state of the art, which can pave the way for future generations of quantum networks. Note that in special cases where the total loss per unit of length is higher than that of the fiber loss, e.g. in passive optical networks with high splitting losses, MA-MDI-QKD offers rate advantages at shorter distances \cite{Norbert}. This could perhaps be the first realistic scenario in which quantum memories, with all their known practical limitations, can be used to offer a tangible benefit. 

\section*{Acknowledgments}
We would like to thank Dr. Koji Azuma for fruitful discussions.
This work was partly funded by the UK's EPSRC Grant EP/M013472/1 and EPSRC
Grant EP/M506951/1.

\appendix

\section{MDI-QKD with imperfect memories}

In this Appendix we explain the general procedure to derive the terms
in Eq. \eqref{eq:key_rate} for the proposed setups in Sec.~\ref{Sec:withNV}. We
consider path loss, given by $e^{-L/L_{\mathrm{at}t}}$ for a distance
$L$ and a channel attenuation length $L_{\mathrm{att}}$, quantum
efficiency $\eta_{D}$, and dark count per pulse $d_{c}$, assuming
that no eavesdropper is present. We also account for memory decoherence,
modeled by a depolarizing channel, which maps an initial state $\rho_{QM}\left(0\right)$
to

\begin{equation}
\rho_{QM}\left(t\right)=p\rho_{QM}\left(0\right)+\left(1-p\right)I/\mathrm{dim\left(\rho_{QM}\left(0\right)\right),}\label{eq:dec_d_mat}
\end{equation}
after a $t$-long period of decoherence, where $I$ is the identity
operator, $p=e^{-t/T_{2}}$ , and $T_{2}$ is the coherence time of
the NV center. This model properly captures the decoherence effect
in an NV center \cite{NV2}.

We first calculate $P_{A}$ and $P_{B}$ by finding the probability
of a successful side-BSM when Alice and Bob use the $Z$ basis for
encoding their bits. This can be done by modeling all the lossy elements
in each leg of Fig. \ref{fig:mem_assisted_setups}(b) by beam splitters
and then simplifying the model by techniques shown in Fig. \ref{fig:BSM_pol}(c).
The resulting butterfly module has been analyzed for relevant input
states in Appendices A and B of \cite{IEEE1}. Here, we avoid duplicating
the same results and simply use them to find the success probabilities
$P_{A}$ and $P_{B}$ as well as the resulting state for Alice and
Bob's QMs after a successful side-BSM.

The second step in our key rate analysis is to derive the error and
yield terms corresponding to the middle BSM. For this, we need to
account for the decoherence in one memory while it waits for the other
memory to be loaded. The decoherence effect can be modeled by using
Eq. \eqref{eq:dec_d_mat} at $p=\exp\left(-|N_{A}-N_{B}|T/T_{2}\right)$,
where $N_{A}$ and $N_{B}$ represent the round at which Alice and
Bob's QMs are, respectively, loaded. $N_{A}$ and $N_{B}$ follow
a geometric distribution with success probabilities $P_{A}$ and $P_{B}$,
respectively. The derivation of yield and QBER terms have been fully
detailed for a dephasing channel in \cite{panayi}. Here, we modify
the analysis in Appendix D of \cite{panayi} to replace the dephasing
channel with the depolarizing channel used here, and carry out the
same calculations as required by Eqs. (3.3) and (3.7) in \cite{panayi}.
The derivations are cumbersome, but with a combination of results
in \cite{IEEE1} and \cite{panayi}, one can find all relevant terms
in Eq. \eqref{eq:key_rate} as a function of the system parameters.
For brevity, the full derivation is left to the reader.

\section*{References}

\bibliographystyle{apsrev4-1}
\bibliography{bib1}

\end{document}